%% file: AAAI26-CameraReady.tex
\documentclass[letterpaper]{article} 
\usepackage{aaai2026}  
\usepackage{times}  
\usepackage{helvet}  
\usepackage{courier}  
\usepackage[hyphens]{url}  
\usepackage{graphicx} 
\usepackage{natbib}  
\usepackage{caption} 
\usepackage{algorithm}
\usepackage{algorithmic}

\usepackage{booktabs}       
\usepackage{amsfonts}       
\usepackage{nicefrac}       
\usepackage{microtype}      
\usepackage{xcolor}         
\usepackage{amsmath}
\usepackage{amssymb}
\usepackage{graphicx}
\usepackage{hhline}
\usepackage{subfig}
\usepackage{makecell}
\usepackage{siunitx}
\usepackage{pifont}
\usepackage{multirow}
\usepackage{tabularx}

\newcommand{\ie}[0]{\textit{i.e.}}
\newcommand{\eg}[0]{\textit{e.g.}}

\usepackage{newfloat}
\usepackage{listings}
\DeclareCaptionStyle{ruled}{labelfont=normalfont,labelsep=colon,strut=off} 
\floatstyle{ruled}
\newfloat{listing}{tb}{lst}{}
\floatname{listing}{Listing}
\title{Force-Aware 3D Contact Modeling for Stable Grasp Generation}
\author{
    Zhuo Chen\textsuperscript{\rm 1},
    Zhongqun Zhang\textsuperscript{\rm 1,2}\thanks{Zhongqun Zhang is the corresponding author.},
    Yihua Cheng\textsuperscript{\rm 1},
    Ale\v{s} Leonardis\textsuperscript{\rm 1},
    Hyung Jin Chang\textsuperscript{\rm 1}
}
\affiliations{
    \textsuperscript{\rm 1}University of Birmingham, 
    \textsuperscript{\rm 2}College of Software, Nankai University \\
    zxc417@student.bham.ac.uk, zhangzhongqun@nankai.edu.cn, \{y.cheng.2,a.leonardis,h.j.chang\}@bham.ac.uk, 

}

\usepackage{bibentry}

\begin{document}

\maketitle

\begin{abstract}
    Contact-based grasp generation plays a crucial role in various applications. Recent methods typically focus on the geometric structure of objects, producing grasps with diverse hand poses and plausible contact points. However, these approaches often overlook the physical attributes of the grasp, specifically the contact force, leading to reduced stability of the grasp. In this paper, we focus on stable grasp generation using explicit contact force predictions. First, we define a force-aware contact representation by transforming the normal force value into discrete levels and encoding it using a one-hot vector. Next, we introduce force-aware stability constraints. We define the stability problem as an acceleration minimization task and explicitly relate stability with contact geometry by formulating the underlying physical constraints. Finally, we present a pose optimizer that systematically integrates our contact representation and stability constraints to enable stable grasp generation. We show that these constraints can help identify key contact points for stability which provide effective initialization and guidance for optimization towards a stable grasp. Experiments are carried out on two public benchmarks, showing that our method brings about 20\% improvement in stability metrics and adapts well to novel objects. Code is available at project webpage: \url{https://chzh9311.github.io/force-aware-grasp-project/}.
\end{abstract}


\input{contents/introduction}

\input{contents/related_work}

\input{contents/method}

\input{contents/experiment}

\input{contents/conclusion}

\input{contents/acknowledgement}

\bibliography{aaai2026}


\end{document}

%% file: contents/introduction.tex
\section{Introduction}

Hand grasp synthesis aims at generating plausible grasping pose of the human hand with an object template \cite{jiang2021graspTTA,karunratanakul2021halo,liu2023contactgen, MACS2024, lee2024interhandgen,zuo2024graspdiff,xu2024dexterous,zhang2025single}. It has recently drawn increasing attention due to its wide applications in AR/VR \cite{grady2024pressurevision++}, robotics \cite{Li2024ShapeGraspZT,zhong2025dexgrasp}, and human-computer interaction \cite{guo2021human}. Previous works have already extensively studied the geometric properties of Hand-Object Interaction (HOI), where contact modeling plays an important role \cite{jiang2021graspTTA,zhou2022toch,grady2021contactopt,zhang2024nl2contact}. Since hand-object contact is very sensitive to noise in hand pose, existing studies prefer to use contacts as geometric constraints to generate hand pose and perform well in plausibility, controllability \cite{zhang2024nl2contact} and diversity \cite{liu2023contactgen}.

\begin{figure*}[t]
    \centering
    \includegraphics[width=.88\linewidth]{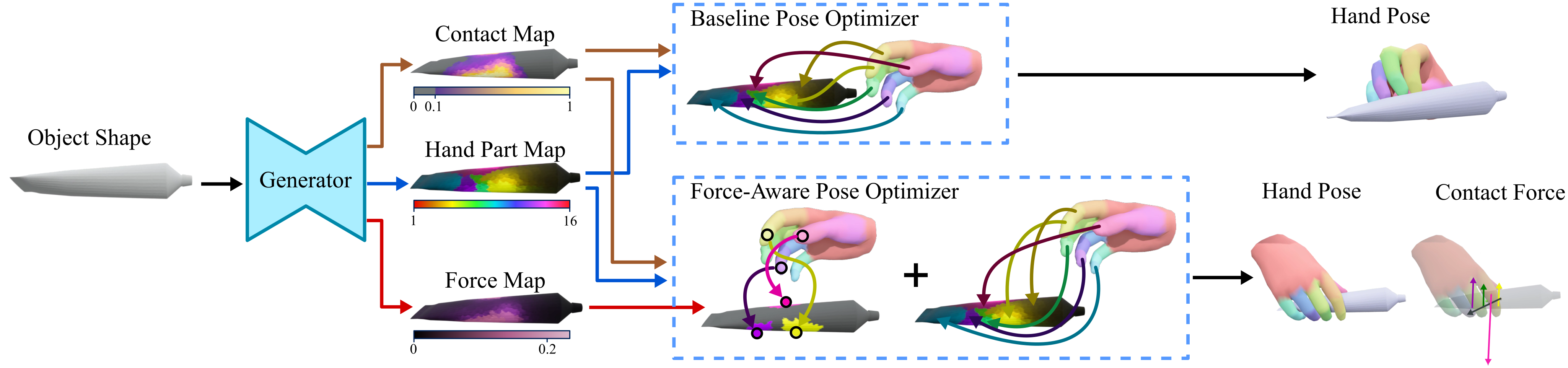}
    \caption{The difference between previous methods (without force predictions) and ours (with force predictions). Force information guides the optimization by identifying a few keypoints to ensure stability.}
    \label{fig:teaser}
\end{figure*}

However, the stability of grasp synthesis has been much less explored. Although some recent studies try to model physical properties in HOI, they are not yet properly related to contact geometry. On the one hand, force-agnostic stability modeling is hard to integrate with contacts. Some existing methods \cite{liu2022forceclosure, luo2024physics} address stability by setting equilibrium constraints with both forces and contact points as unknowns, which requires strong assumptions like zero-friction to obtain solvable constraints, leading to suboptimal results. Others directly use a simulator \cite{zhao2022stability, yuan2023physdiff} in inference, but iterative simulations are too complicated to directly incorporate with contact-based pose refinement. On the other hand, existing studies the relation between forces and visual clues or object mass instead of contact geometry, and are only limited to specific object shapes, \eg, boards \cite{grady2022pressurevision, grady2024pressurevision++,zhao2024egopressure} or cubes \cite{murtaza2023qmcube}. The relationship between grasping geometry and contact forces on general objects remains unexplored.

Meanwhile, we find that if the contact forces can be predicted, then modeling stability in contact-based grasp synthesis will be much easier. This is since stability can be measured as the object acceleration and is related to contact forces and points via physical laws. With the forces known, it is possible to identify the most important contact keypoints for static equilibrium without ignoring gravity and frictions. These keypoints provide valuable guides in grasp synthesis. They are simple geometric targets that a hand model \cite{romero2022embodied} can easily fit, but they also contain important physics information so that successfully contacting these points largely ensures stability, as shown in Fig.~\ref{fig:teaser}.

Thus, we propose to explicitly integrate forces into 3D contact modeling. Specifically, we present a novel force-aware contact representation, based on which we construct a pipeline with a contact generator and a pose optimizer following previous contact-based pose synthesis methods \cite{grady2021contactopt, liu2023contactgen, zuo2024graspdiff}. In the framework, we first extend the contact representation to include contact likelihood, hand part label, and contact normal force value for each sampled point on the object surface. The generator is then trained to generate all three maps. In the optimization stage, we model stability as the minimization of accelerations based on which a few key contact points are recognized. Then a strong initialization is performed by fitting the hand model \cite{romero2022embodied} to the keypoints. The pose optimizer then tries to fit the hand model to the keypoints and contact maps, leading to a stable grasp.


To train our generator, we need ground-truth force labels. However, existing HOI datasets with force labels \cite{grady2022pressurevision,grady2024pressurevision++,zhao2024egopressure,murtaza2023qmcube} are all limited to specific object shapes. Thus, we develop an automatic labeling pipeline using a simulator \cite{todorov2012mujoco} to obtain the force label on general objects from existing HOI datasets \cite{GRAB:2020}. The procedure searches for the optimal simulation parameters to restore the stability of ground truth data and output faithful force labels.

Finally, we test the generated grasp quality on two public benchmarks: GRAB \cite{GRAB:2020} and HO3Dv2 \cite{hampali2020honnotate}. The results show that our method brings over 30\% improvement in stability criteria on both in-domain and out-of-domain objects without affecting plausibility and diversity. It shows that our method effectively addresses stable grasp synthesis and adapts well to unseen objects. 

Our contributions can be summarized as follows:

\begin{enumerate}
    \item We formally model the relation between stability and contact points using force. We show that by optimizing accelerations, stability can be formulated as a solvable optimization of contact points and predicted contact forces;
    \item We propose a novel pipeline for stable grasp synthesis. The pipeline is built upon force-aware contact representation, with a generator and optimizer. The former relation is utilized to identify keypoints to guide grasp generation;
    \item We conduct experiments on two public benchmarks, showing that our method achieves state-of-the-art stability and excellent generality to novel objects.
\end{enumerate}


%% file: contents/related_work.tex
\section{Related Work}
\label{sec:related_force_model}

\subsection{Contact-Based Grasp Synthesis}
As physical plausibility can be affected by slight changes in hand pose \cite{grady2021contactopt}, modeling hand-object contact has been a key to synthesizing plausible and stable grasps. Some methods propose to generate hand poses directly, and refine the poses using predicted contact maps \cite{jiang2021graspTTA, GRAB:2020, karunratanakul2021halo,taheri2022goal}. Usually, the generator is a cVAE \cite{sohn2015learning} and the refiner tries to fit MANO \cite{romero2022embodied} model to the predicted contacts by training a separate model \cite{GRAB:2020} or optimizing iteratively \cite{jiang2021graspTTA, taheri2022goal}. Other methods generally follow a two-stage pipeline to generate contact maps then optimize the hand pose to fit the contact \cite{grady2021contactopt}. ContactGen \cite{liu2023contactgen} extends the contact representation to include hand part labels and orientations of the hand parts for better diversity, and GraspDiff \cite{zuo2024graspdiff} integrates the optimization process into Diffusion \cite{ho2020ddpm} inference. Great improvements in plausibility and diversity have been observed, but stability is largely ignored because of the indirect relations between physical properties and geometry. In contrast, our research provides a novel perspective to bridge the gap via contact forces, which greatly improves the stability while keeping the advantages of contact-based synthesis.



\subsection{Modeling Grasp Stability}
Grasping stability has long been an important criterion of grasp generation \cite{jiang2021graspTTA}. It offers guidance to increase grasp synthesis quality \cite{liu2022forceclosure, yuan2023physdiff}. For stability-aware pose generation, physics engines \cite{coumans2021pybullet,todorov2012mujoco,makoviychuk2021isaac} are commonly used in inference by projecting the hand or human pose to physically plausible space via a mimic agent learnt by RL \cite{yuan2023physdiff}. Though being more stable, the relations are not modeled explicitly, and the process is hard to integrate with contact modeling. So some other studies try to model stability according to equilibrium-related theories, like friction cone constraints  \cite{luo2024physics} or simplified force closure \cite{nguyen1988constructing} conditions \cite{liu2022forceclosure}. One major challenge is the high dimensionality with both forces and contact points as unknowns, thus strong assumptions like zero-friction are applied, resulting in suboptimal results. Therefore, we propose to predict forces as part of the known variables, which directly solves this issue, allowing more precise equilibrium relations to be modeled in the synthesis process.


\begin{figure*}[t]
    \centering
    \includegraphics[width=1.0\linewidth]{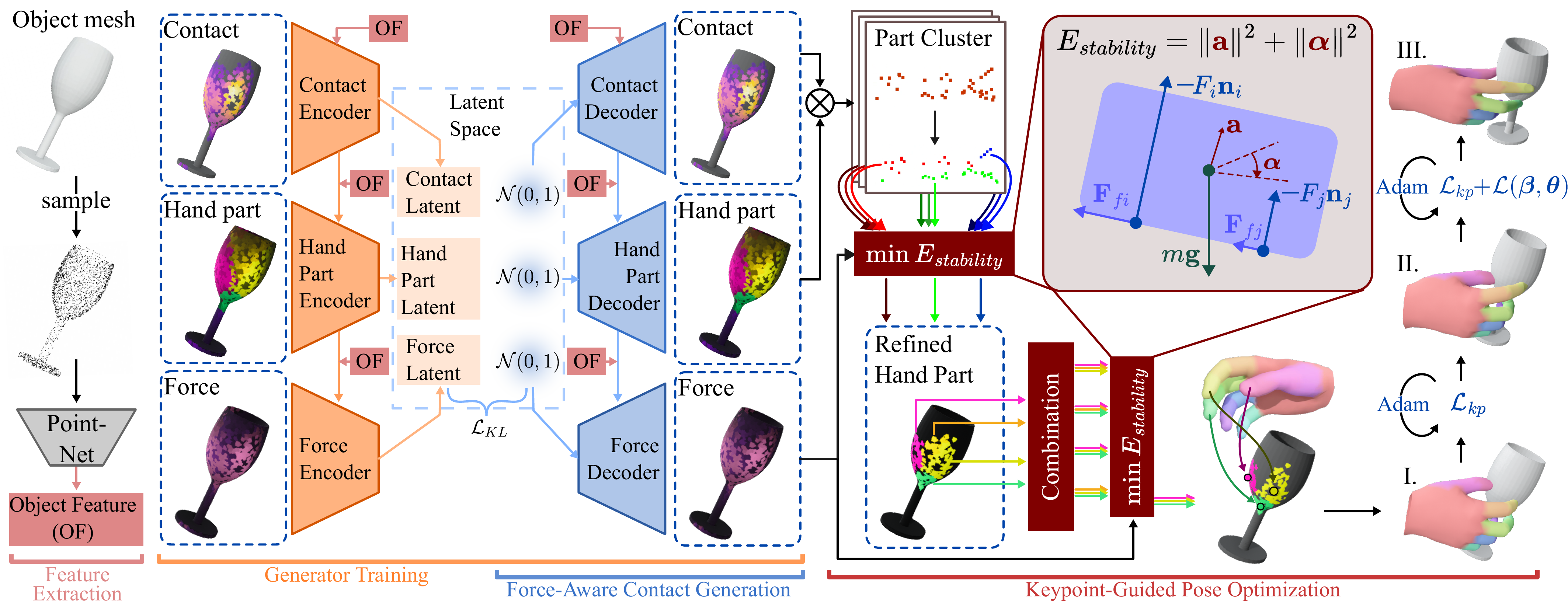}
     \caption{Our framework consists of a generator and an optimizer. The object features (OF) are obtained by processing sampled object points using PointNet++. Then the generator consisting of 3 successive cVAEs takes OF as an input condition. It generates force-aware contacts including the contact normal force value, which are then used to recognize important keypoints for equilibrium. In optimization stage, I. and II. are two-stage initialization based only on keypoints, and III. optimizes MANO parameters $\boldsymbol{\beta}, \boldsymbol{\theta}$ while also being guided by the keypoints to get the stable grasp pose.}
    \label{fig:framework}
\end{figure*}



%% file: contents/method.tex
\section{Method}


In this section, we first describe our strategy for acquiring contact force labels for model training. Then we define our force-aware contact representation and the stability constraints, which form the core of our approach. Finally, we describe our model, which incorporates the representation and constraint to enable stable grasp synthesis. The framework is shown in Fig.~\ref{fig:framework}.


\subsection{Automatic Contact Force Labeling}
Existing datasets typically rely on physical sensors to acquire force labels, which limits the data to plane-related objects, such as board-based \cite{grady2022pressurevision, grady2024pressurevision++, zhao2024egopressure} or cube-based \cite{murtaza2023qmcube} interactions. In this paper, we obtain contact force labels using a physics simulator~\cite{todorov2012mujoco}, based on hand and object meshes that can be derived from existing hand-object interaction datasets such as GRAB~\cite{GRAB:2020}.

To allow enough tolerance to the unavoidable errors in GT data and approximately modeling soft tissues, we propose to do \textit{dynamic parameter searching}. First, we set the object free and take the contact forces when the object achieves the least acceleration within a certain displacement threshold. Then, we search for the hyperparameter setting resulting in the least displacement as the optimal parameters for the current sample, increasing the stable rate (displacement $<$ \SI{5}{\centi\metre}) to 65.4\%. Please refer to supplementary for details in labeling procedure and data statistics.

The final label is in point-contact form. One point from the $j^{\text{th}}$ contact area is determined by the simulator as the contact point $\mathbf{c}_j\in\mathbb{R}^3$. The label for this point also includes normal force value $N_j$ (supportive force), lateral force (friction), and which hand part provides the force (hand part label).

\subsection{Force-Aware Contact Representation}

We use three hierarchical feature maps for force-aware contact representation. For each sampled point from the object surface, three scalars are assigned to respectively indicate the contact likelihood $C \in[0, 1]$, the hand part label in contact $P\in[1, 16]$, and the normal contact force value $F\in\mathbb{R}^+$.

With the hand and object meshes, the contact likelihood of each point is defined using the capsule distance $d$ in \cite{grady2021contactopt} between the object point and the nearest vertex of the hand mesh: $C=\min\{1/d, 1\}$. The part label is defined as the hand part of the nearest vertex, where the hand is partitioned according to blend-skinning weights of MANO. Hands in Fig.~\ref{fig:framework} show the hand parts with different colors.

For contact force, we propose to only use the normal force values in our representation. On the one hand, including frictions brings two extra dimensions for each contact point and makes it harder to predict. On the other hand, the friction can be inferred from the equilibrium constraints using the normal force and is therefore unnecessary to predict. Finally, the force prediction is a scalar $F_i \geq 0$ for the $i^{\text{th}}$ sampled point. And because the simulator labels forces on discrete points not continuous areas, we need to spread the force uniformly on the affinity object point set $A_j$, which includes all valid contact points whose nearest label point is $\mathbf{c}_j$, \ie, $\forall i \in A_j, F_i = N_j/|A_j|$.





We then propose to represent the force values using one-hot vector. This is since classification-based methods empirically outperforms regression-based ones \cite{li2021human}. To build the discrete force distribution, we define a binning scheme over $F$ using $s$ bins. The first bin captures the probability of zero force ($F=0$), while the remaining $s-1$ bins cover the positive force range. To define the bin boundaries over the non-zero range, we assume the logarithm of the force magnitude, $\log F$, follows a normal distribution with mean $\mu_{\log F}$ and standard deviation $\sigma_{\log F}$. We uniformly divide the interval $[\mu_{\log F} - 3\sigma_{\log F}, \mu_{\log F} + 3\sigma_{\log F}]$ into $s-2$ intervals. Set $l_{F1} = 0, l_{F(s+1)}=\infty$, and
\[
\forall\, 2 \leq i \leq s,\; l_{Fi} = \exp\left\{\mu_{\log F} + \left(\frac{6(i-2)}{s-2} - 3\right)\sigma_{\log F}\right\}.
\]
This construction ensures that the non-zero force range is covered by $s-2$ bins uniformly distributed in log-space, accommodating the heavy-tailed nature of the force distribution.

The vectorized force is defined by Eq.\eqref{eq:f2v}, and the transformation from the one-hot vector back to a scalar force value is performed using the soft-argmax operation with a temperature parameter $t = 0.02$, as shown in Eq.\eqref{eq:v2f}.

    \begin{equation}\label{eq:f2v}
        \mathbf{v}_{Fi}=\begin{cases}
            1, & \text{if}\; l_{Fi} \leq F_i <l_{Fi+1}; \\
            0, & \text{otherwise}.
        \end{cases}
    \end{equation}
    
    \begin{equation}\label{eq:v2f}
        \hat{F} = \frac{\sum_{i=1}^s \frac{l_{Fi}+l_{Fi+1}}{2}\exp\{\mathbf{v}_{Fi}/t\}}{\sum_{i=1}^s \exp\{\mathbf{v}_{Fi}/t\}}.
    \end{equation}

\subsection{Modeling Stability with Forces}


The definition of stability can vary with applications, yet the ultimate goal is always to keep the object still. Formally, it means both the translational acceleration $\mathbf{a}\in\mathbb{R}^3$ and rotational acceleration $\boldsymbol{\alpha}\in\mathbb{R}^3$ are 0. Then our basic idea is to measure stability using the norm of accelerations.


\begin{figure}
    \centering
    \includegraphics[width=0.6\linewidth]{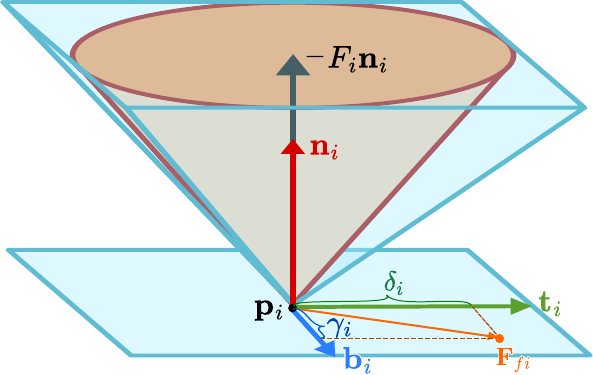}
    \caption{The friction cone (red) and its approximation (blue) that is used in formulating the stability energy function}
    \vspace{-1ex}
    \label{fig:friction_cone}
\end{figure}

With the predicted contact forces, it is possible to calculate the exact accelerations via physical laws. Denote the center of mass of the object as $\mathbf{p}_{CoM}$. Then formally, according to \textit{Newton's Second Law} and \textit{the Theory of Angular Momentum}:
\begin{eqnarray}
    \sum_{i=1}^n(-F_i\mathbf{n}_i + \mathbf{F}_{fi}) + m\mathbf{g} &=& m\mathbf{a}, \label{eq:acc}\\
    \sum_{i=1}^n (\mathbf{p}_i-\mathbf{p}_{CoM}) \times (-F_i\mathbf{n}_i + \mathbf{F}_{fi}) &=& I\boldsymbol{\alpha} \label{eq:momentum},
\end{eqnarray}
where $\mathbf{g}$ is the gravitational acceleration, $\mathbf{F}_{fi} = \mu F_i(\gamma_i\mathbf{b}_i + \delta_i\mathbf{t}_i)$ is the friction, $-1 \leq \gamma_i, \delta_i \leq 1$. $(\mathbf{b}_i,\mathbf{t}_i,\mathbf{n}_i)$ forms a unit orthogonal basis, and $\mathbf{n}_i$ is the surface normal. $\mu$ is the friction coefficient empirically set to 1. It is a common linear approximation for the friction cone \cite{hu2022physical} as shown in Fig.~\ref{fig:friction_cone}. The object mass $m$ is set to \SI{1}{\kilo\gram} for normalization, and the moment of inertia $I$ is approximated by regarding the 
object as a ball: $I=0.4m\max_i\|\mathbf{p}_i-\mathbf{p}_{CoM}\|^2$.

Eq.~\eqref{eq:acc} and \eqref{eq:momentum} can be further formulated in a bilinear form as
\begin{equation} \label{eq:mat_acc}
    N\mathbf{F} + \mu B(\boldsymbol{\gamma}\circ\mathbf{F}) + \mu T(\boldsymbol{\delta}\circ \mathbf{F}) + \begin{bmatrix}\mathbf{g} & 0\end{bmatrix}^T = \begin{bmatrix}\mathbf{a} & \boldsymbol{\alpha}\end{bmatrix}^T,
\end{equation}
where $N = [N_m/m, N_t/I]^T$ and $N_m=[\mathbf{n}_1, \mathbf{n}_2, \dots, \mathbf{n}_n]^T$, $N_t = [(\mathbf{p}_1-\mathbf{p}_{CoM})\times\mathbf{n}_1, (\mathbf{p}_2-\mathbf{p}_{CoM})\times\mathbf{n}_2, \dots, (\mathbf{p}_n-\mathbf{p}_{CoM})\times\mathbf{n}_n]^T$. $B$ and $T$ follow this definition. ``$\circ$'' is the Hadamard product. $n$ is the point number and $\mathbf{F}=[F_1, F_2, \dots, F_n]^T$ represents the force map in representation.

In Eq.~\eqref{eq:mat_acc}, only if $\mathbf{F}$ is known can the acceleration vector $[\mathbf{a}, \boldsymbol{\alpha}]^T$ be linearly related to $\boldsymbol{\gamma}$ and $\boldsymbol{\delta}$. Then we define the stability energy as the norm of accelerations, which is formulated as a constrained quadratic optimization problem and can be solved by quadratic programming: 
\begin{eqnarray}
    &E_{stability}&\hspace{-1em}(\{\mathbf{F}_i\}_{i=1}^n, \{\mathbf{p}_i\}_{i=1}^n) = \min_{\boldsymbol{\gamma}, \boldsymbol{\delta}} \left\{\|\mathbf{a}\|^2 + \|\boldsymbol{\alpha}\|^2\right\}, \nonumber\\ 
    & &\text{s.t.}\; \forall 1\leq i \leq n,  -1 \leq \gamma_i, \delta_i \leq 1.\label{eq:stability_energy}
\end{eqnarray}

Since quadratic programming is non-differentiable, to integrate stability loss into training, we check the bounds of each element in $\mathbf{a}$ and $\boldsymbol{\alpha}$, and punish the cases when 0 is not inside.  Use $\text{abs}(\cdot)$ to represent the element-wise absolute operator, and $B_{i\cdot}$ is the $i^{\text{th}}$ row of $B$, then $-\text{abs}(B_{i\cdot})\mathbf{F} \leq B_{i\cdot} (\boldsymbol{\gamma}\circ\mathbf{F}) \leq \text{abs}(B_{i\cdot})\mathbf{F}$. The same goes for $T$. Thus, we can derive a differentiable loss term by comparing the upper or lower bounds of each element in Eq.~\eqref{eq:mat_acc} to 0 if 0 is outside the range. Using the notation $F_{fric} = \mu\;\text{abs}(B) + \mu\;\text{abs}(T)$, the loss term can be written as

\begin{equation}
\label{eq:stability_loss}
\begin{split}
    \mathcal{L}_{stability} &= \mathbf{1}^T\max\left\{\left(N - F_{fric}\right)\mathbf{F} + \begin{bmatrix}\mathbf{g} & 0\end{bmatrix}^T, 0 \right\}\\
     &- \mathbf{1}^T\min\left\{\left(N + F_{fric}\right)\mathbf{F} + \begin{bmatrix}\mathbf{g} & 0\end{bmatrix}^T, 0\right\}
\end{split}
\end{equation}


\subsection{Force-Aware Stable Grasp Synthesis}


\textbf{Generator Structure and Training Losses}. The generator consists of 3 successive cVAEs, similar to \cite{liu2023contactgen}. The positions and normals of all sampled object points are processed by PointNet++ \cite{qi2017pointnetplusplus}, and the features are fed to all encoders and decoders as a part of the condition. The latent output of the former forms the other part.

The model is trained using reconstruction loss, KL-divergence loss, and stability loss. The reconstruction loss $\mathcal{L}_{rec} = \mathcal{L}_{c} + \mathcal{L}_{part} + \mathcal{L}_{force}$. The hand part map loss $\mathcal{L}_{part}=CE(\mathbf{P}, \hat{\mathbf{P}})$ is cross entropy loss, while contact loss $\mathcal{L}_{c}$ and force loss $\mathcal{L}_{force}$ are both L1 loss. KL-divergence $\mathcal{L}_{KL}$ on latent codes is also enforced following the training process of cVAE \cite{sohn2015learning}. Note that in Eq.~\eqref{eq:stability_loss} we substitute $\mathbf{F}$ with $\hat{\mathbf{F}} \circ \hat{\mathbf{C}}$ to also regularize contact. Considering not all samples are stable, and we prefer stable samples to unstable ones, the total loss is weighted according to the displacement $d$ in simulative labeling by $w=\min\{d_{th}/d, 1\}$ where $d_{th}=\SI{5}{\centi\metre}$. With preset weights, the final loss is

\begin{equation}
    \mathcal{L}_{cVAE} = w\left(w_{rec}\mathcal{L}_{rec} + w_{KL}\mathcal{L}_{KL} + w_{stability}\mathcal{L}_{stability}\right).
\end{equation}

\noindent\textbf{Keypoint-Guided Stable Pose Optimization}. The contact forces can provide guidance on which contact points are more important than others, but simply using the force values as weights results in suboptimal results (See Tab.~\ref{tab:ablation} for details), since the optimization based on MANO \cite{romero2022embodied} is still non-convex. However, a better pose initialization is still possible, so the optimizer is less likely to get stuck in local minima. Previous studies apply random restarts \cite{grady2021contactopt} or iterative optimization \cite{liu2023contactgen}, which are either non-robust or too complex to avoid local minima.

According to the theory of force closure \cite{nguyen1988constructing}, as few as 3 contact points are enough for a stable grasp in 3D. The energy function in Eq.~\eqref{eq:stability_energy} is well suited for selecting the optimal contact points. Therefore, we propose to fit the MANO model to only a few keypoints that are important for object equilibrium as a strong initialization and stability-aware guidance through optimization. 

To prevent the keypoints from falling inside the object (e.g., when the object is slim), the contact points of each part are first clustered and only one cluster is selected to produce the keypoint. We aggregate the supportive forces and contact points of cluster $c$ of hand part $h$ via force-weighted average over all the points as $\mathbf{F}_c^h$ and $\mathbf{p}_c^h$. Suppose $H$ is the set of hand parts that are in touch with the object, then the optimal cluster label is the one with the least stability energy, \ie,


\begin{equation}\label{eq:cluster_select}
    c_0 = \underset{c}{\text{argmin}} E_{stability}\left(\{\mathbf{F}_c^h, \mathbf{p}_c^h\}\cup \{\mathbf{F}_i, \mathbf{p}_i\}_{i\in H, i\neq h}\right).
\end{equation}

Then we assign $\mathbf{F}_h=\mathbf{F}_{c_0}^h,  \mathbf{p}_h=\mathbf{p}_{c_0}^h$ as the normal force and contact center of part $p$ and iterate Eq.~(\ref{eq:cluster_select}) over all contact parts. 

Next, we want to select $n_{kp}$ keypoints from the contact centers. Similarly, the optimal keypoint combination is the one with the least $E_{stability}$. In experiments, we set $n_{kp}=3$ because it is the least possible number of contact points that forms a force closure, and a smaller number of target points means more traceable optimization. Thus, the optimal combination part labels is
\begin{equation}
    H_0 = \underset{\substack{H_c\subset H \\ |H_c|=n_{kp}}}{\text{argmin}} E_{stability}\left(\{\mathbf{F}_h\}_{h\in H_c}, \{\mathbf{p}_h\}_{h\in H_c}\right).
\end{equation}
Then the target point of the hand part is obtained by moving the contact points along the surface normal for $r=\SI{5}{m\metre}$, \ie, $\mathbf{q}_i=\mathbf{p}_i + r\mathbf{n}_i$, roughly the finger radius.

\noindent\textbf{Two-stage Initialization}. Although fitting the hand joints to only 3 keypoints seems simple, the problem is still hard to solve analytically due to the non-linearity of the MANO model. Iterative optimization is an intuitive solution, but it is still possible to get stuck in local minima, and the initialization problem remains unsolved. Instead, in the first stage, we fit the average pose to the keypoints. Thus, the hand part centers are relatively fixed, and the problem then becomes point cloud registration. Suppose $\{\mathbf{p}^{hand}_h\}_{h\in H}$ are the hand part centers defined as the average position of connected joints, then the problem is registering $\{\mathbf{p}^{hand}_h\}_{h\in H_0}$ to $\{\mathbf{p}_h\}_{h \in H_0}$, which can be solved analytically via corresponding point set registration \cite{correspondingPCReg}, where $\mathcal{L}_{kp} = \sum_{h\in H_0}\|\mathbf{p}_h - \mathbf{p}^{hand}_h\|^2$ is minimized.

In the second stage, we optimize the hand pose parameters $\boldsymbol{\theta}$ together with the global pose to further minimize $\mathcal{L}_{kp}$ via iterative optimization. With the former global registration, such optimization is more likely to find the optimal solution and provides a stable initialization.

\noindent\textbf{Keypoint-Guided Optimization}. Based on the initialization, the next step would be to search for the optimal grasping pose around, where more contacts and less penetration should also be considered alongside stability. Inspired by \cite{liu2023contactgen}, we use partitioned SDF to calculate the hand-object distance, and calculate the current contact map $\hat{\mathbf{C}}$ following \cite{grady2021contactopt}. The penetration loss $\mathcal{L}_{pene}$ follows the definition of \cite{liu2023contactgen}. The keypoint loss is kept in this stage as $\mathcal{L}_{kp}$. Together, regularization losses $\mathcal{L}_{reg}=\|\boldsymbol{\beta}\|^2+\|\boldsymbol{\theta}\|^2$ are added to reduce implausible poses. Then the target of the optimization is as follows and is solved by an optimizer (we use Adam \cite{Adam} in our experiment).
\begin{equation}\label{eq:optimizer_target}
    \min_{\boldsymbol{\beta}, \boldsymbol{\theta}}\mathcal{L}_{optim} = w_{kp} \mathcal{L}_{kp} + w_{c}\mathcal{L}_{c} + w_{pene}\mathcal{L}_{pene} + w_{reg}\mathcal{L}_{reg}.
\end{equation}


%% file: contents/experiment.tex
\section{Experiment}

\subsection{Datasets and Metrics}

\textbf{Datasets and Baselines}. GRAB \cite{GRAB:2020} is used as our training and test set due to its precise hand shapes resulting from motion captures. The 10 objects from HO3Dv2 \cite{hampali2020honnotate} are used to test the out-of-domain adaptivity of the method. During test, the input object is randomly rotated while the gravity direction keeps unchanged. Please refer to the supplementary for training details.

For fair comparison, we compare our method to those trained only with GRAB on GRAB benchmark. On HO3D benchmark, however, all methods are \textit{not} trained on HO3D dataset, including methods from GRAB benchmark and GF \cite{karunratanakul2020grasping} and GraspTTA \cite{jiang2021graspTTA} which are trained on ObMan \cite{hasson19_obman} dataset.


\begin{figure*}[t]
\begin{minipage}[c]{.77\textwidth}
    \centering
    \includegraphics[width=\linewidth]{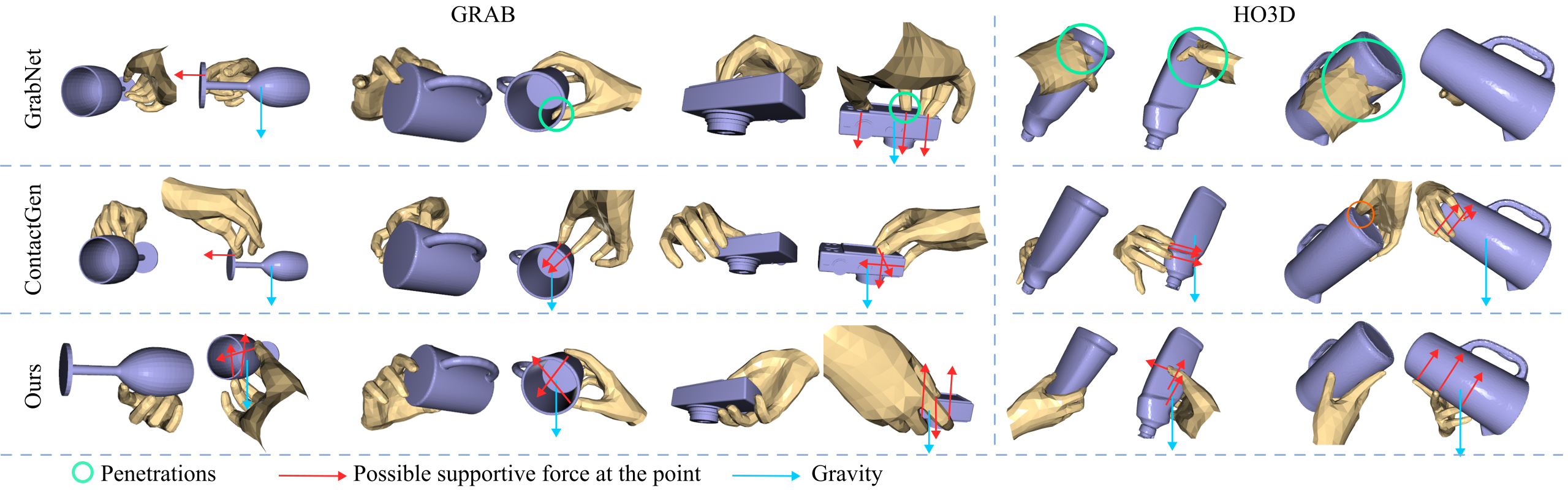}
    \captionof{figure}{Grasp Samples from both datasets. Each sample is shown in two views with highlights of penetrations, possible supportive forces at the contact points, and gravity labels.}
    \label{fig:quant_samples}
\end{minipage} \hfill
\begin{minipage}[c]{.21\textwidth}
    \centering
    \includegraphics[width=\linewidth]{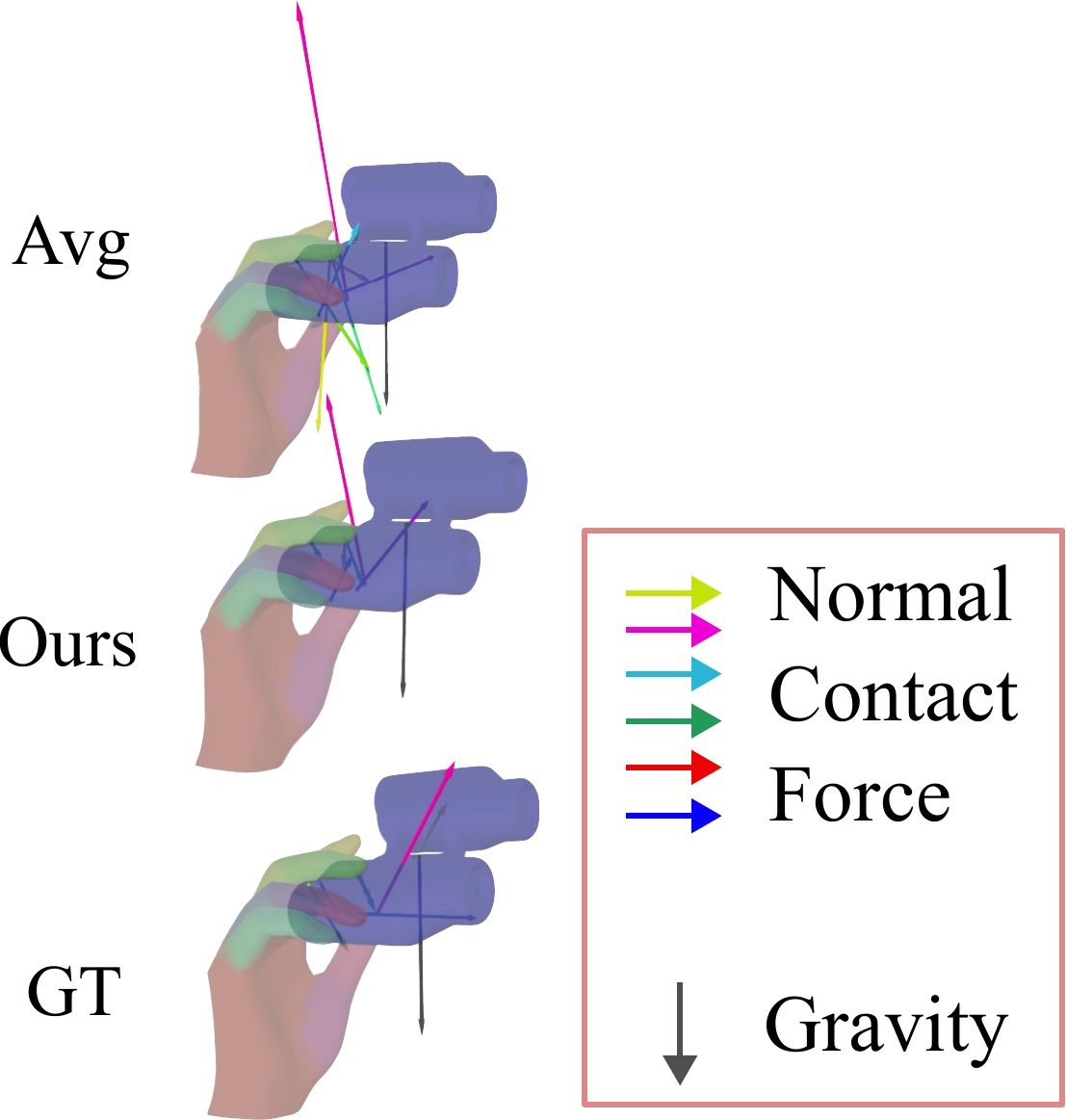}
    \captionof{figure}{Normal contact forces by our method, avg. predictor, and GT}\label{fig:force_pred}
\end{minipage}
\end{figure*}

\noindent\textbf{Stability Assessment}. Following previous research \cite{karunratanakul2020grasping, jiang2021graspTTA, liu2023contactgen, zuo2024graspdiff}, we use PyBullet \cite{coumans2021pybullet} to simulate the object and the grasping hand and measure the displacement of the object center of mass. \textit{Simulation Disp.} is the average displacement while \textit{Stable Rate} is the proportion of samples with Simulation Disp. less than \SI{2}{\centi\metre}. To more comprehensively measure stability considering both displacements and penetrations, we adopt distinction rate from dexterous robotic grasping research \cite{wangDexGraspNetLargeScaleRobotic2023,luUGGUnifiedGenerative2024}. It is the proportion of samples where the grasp can withstand the gravity of the object in IsaacGym \cite{makoviychuk2021isaac} simulation and the maximum penetration depth is less than \SI{5}{m\metre}. Since human hand labels usually represent deformations with slight penetrations \cite{grady2021contactopt}, this criterion is rather strict for human hands, but it is reasonable by showing the proportion of top-quality grasps. Therefore, we rename it to \textit{distinction rate}.

\noindent\textbf{Plausibility and Diversity Assessment}. We use contact ratio and penetration volume to assess the geometric plausibility. \textit{Contact ratio} is the proportional of samples where the hand and object are in contact (nearest distance $<$ \SI{5}{m\metre}). \textit{Penetration Vol.} measures the intersecting volume of the hand and the object. Diversity is assessed by the \textit{Entropy} and \textit{Cluster Size} of the clustering result of all generated hand poses following \cite{liu2023contactgen, wu2024fastgrasp}.

\subsection{Grasp Synthesis Analysis}

\begin{table}[t]
\setlength{\tabcolsep}{1.6mm}
\centering
\small{\begin{tabular}{@{}lccc@{}}
\toprule
\textbf{Methods} & \thead{Simulation \\ Disp. (\unit{\centi\metre})$\downarrow$} & \thead{Stable ($<$\SI{2}{\centi\metre}) \\ Rate (\%)  $\uparrow$} & \thead{ Distinction \\ Rate (\%) $\uparrow$} \\ \midrule 
\multicolumn{4}{l}{\thead[l]{\hspace{-0.5em}\textbf{Results on GRAB dataset} \cite{GRAB:2020}\\ \vspace{-1.5em}}}\\
\midrule
GrabNet \shortcite{GRAB:2020}   & 0.90           & 91.6   &  9.2     \\
ContactGen \shortcite{liu2023contactgen} & 2.82          & 72.5         & 12.5        \\
FastGrasp \shortcite{wu2024fastgrasp} & 2.04             &  74.2       & 14.2      \\ \midrule
Ours       & \textbf{0.68} \textsuperscript{\textcolor{red}{$\downarrow 24.4\%$}} & \textbf{96.7} \textsuperscript{\textcolor{red}{$\uparrow 5.5\% $}} & \textbf{17.5}\textsuperscript{\textcolor{red}{$\uparrow 23.2\%$}} \\
\midrule \midrule
\multicolumn{4}{l}{\thead[l]{\hspace{-0.5em}\textbf{Results on out-of-domain HO3D dataset} \cite{hampali2020honnotate} \\ \vspace{-1.5em}}}\\ 
\midrule
GF  \shortcite{karunratanakul2020grasping}       & 4.47          & 45.0         & 1.0      \\
GrabNet  \shortcite{GRAB:2020}  & 1.86          & 74.0          &   3.0     \\
GraspTTA  \shortcite{jiang2021graspTTA} & 1.91          & 74.0          & 7.5       \\
ContactGen \shortcite{liu2023contactgen} & 4.62          & 51.0         & 9.0      \\
FastGrasp \shortcite{wu2024fastgrasp} & 3.59          & 51.0         & 1.5       \\\midrule
Ours       & \textbf{1.50}\textsuperscript{\textcolor{red}{$\downarrow 19.4\%$}}  & \textbf{80.5}\textsuperscript{\textcolor{red}{$\uparrow 8.5\%$}} & \textbf{17.0}\textsuperscript{\textcolor{red}{$\uparrow 88.9\%$}} \\ \bottomrule
\end{tabular}
 }
\caption{Quantitative results of stability performance on GRAB and HO3Dv2. `$\uparrow$' after a criterion means the higher the better, while `$\downarrow$' means the opposite. Our method outperforms previous ones by a large margin.}\label{tab:quant_result}
\end{table}

\textbf{Quantitative Analysis on Stability}. The stability performance on both datasets are shown in Tab.~\ref{tab:quant_result}.  On both benchmarks, our method achieves the best performance on all criteria, with the displacements decreased amazingly by 24.4\% and 19.4\%. The stable rate and distinction rate also increase obviously. Since the distinction rate considers both penetrations and stability, it is safe to conclude that our method finds a good balance between the two.

On the HO3D benchmark, since no training samples are seen, previous methods all suffer from an obvious performance drop on all metrics. In contrast, the stable rate and distinction rate of our method do not drop obviously. Since the keypoints are selected with physics-based analytical analysis instead of data-driven methods, our method shows strong adaptivity to general out-of-domain objects.

\noindent\textbf{Analysis on Geometric Plausibility and Diversity}. We report the relative criteria in Tab.~\ref{tab:penetration_diversity}. Our method shows the best contact ratio and comparable diversity. Although the penetration volumes are increased compared to previous methods, the state-of-the-art distinction rate in Tab.~\ref{tab:quant_result} indicates that the increase is acceptable. The good diversity partly results from the change of object pose relative to the gravity direction. With Eq.~\eqref{eq:stability_loss}, our model favors more supportive forces in the anti-gravity direction, thus leading to a gravity-direction-related generation. By changing the gravity, our method can generate diverse and stable grasps.

\noindent\textbf{Qualitative Analysis}. To study the reasons for improvements, we visualize some generated samples from our method and two previous ones in Fig.~\ref{fig:quant_samples}.

First, there is a trend that our method favors hands supporting the object from the bottom, which is preferable as they provide supportive forces against gravity. Fingers tend to touch the object from different directions due to the force-aware choice of keypoints. In contrast, GrabNet tends to produce penetrations (also shown in Tab.~\ref{tab:quant_result}), and ContactGen prefers less contact, usually unstable grasps. Further, we can observe a common failure pattern from previous methods is that the hand ends up in touch with only one side of the object, like the `camera' of GrabNet and `mug' of ContactGen in Fig.~\ref{fig:quant_samples}. These cases mostly initialize the hand on one. Later optimization or refinement tries to lead the fingers through the object but is prevented by penetration losses. This indicates the importance of a stability-aware pose initialization.

\begin{figure}[t]
        \subfloat[GRAB]{
            \includegraphics[width=.48\linewidth]{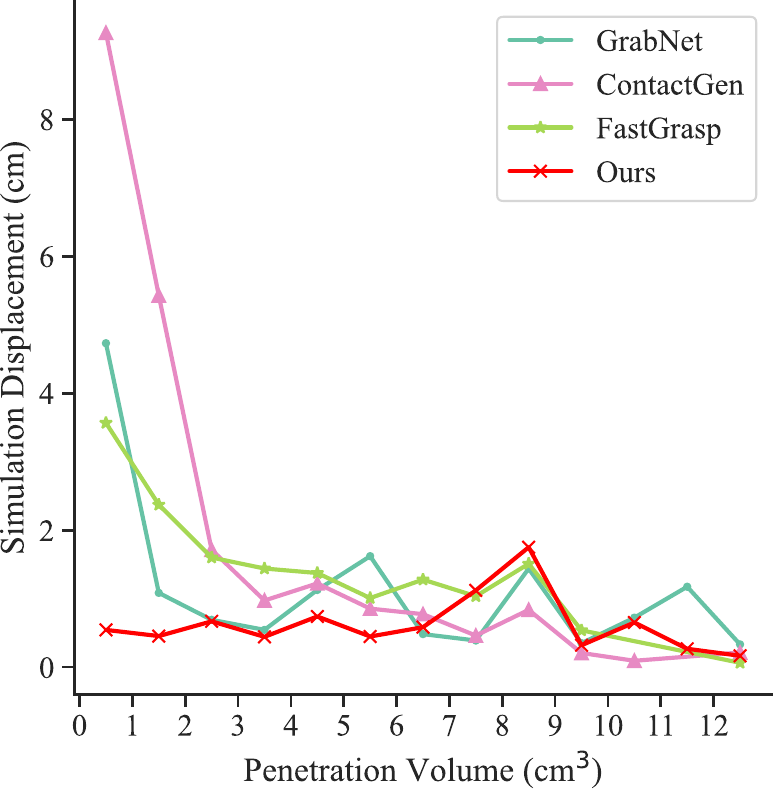}
        }
        \subfloat[HO3D]{
            \includegraphics[width=.48\linewidth]{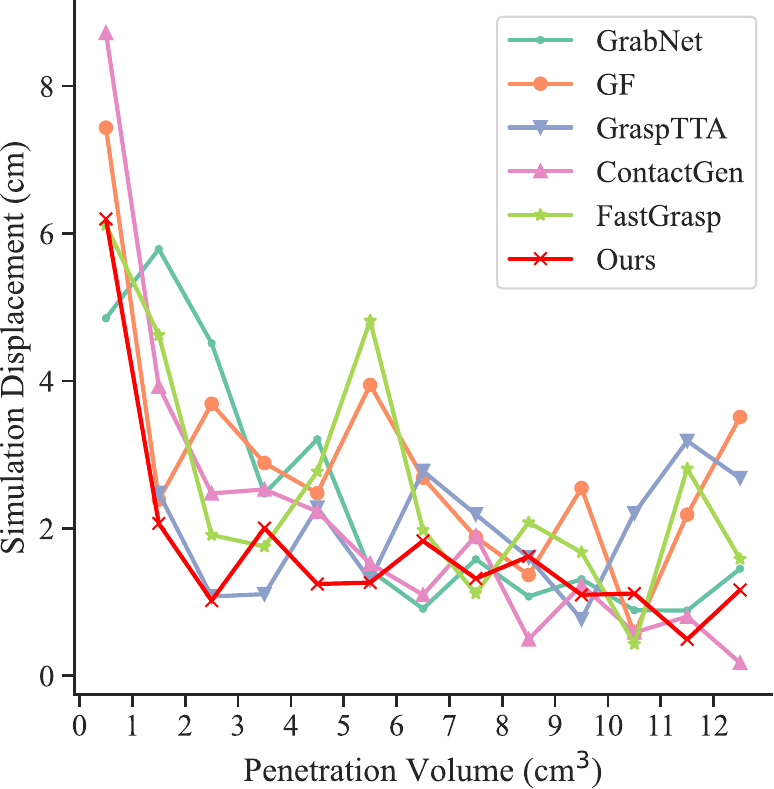}
        }
        \caption{Simulation displacements in different penetration intervals. The curve of our method is in bright red and is the lowest especially for small penetrations.}
    \label{fig:disp-pene}
\end{figure}

\noindent\textbf{Discussion on Stability and Penetration}. Previous research has observed that larger penetrations are usually accompanied by smaller displacements \cite{jiang2021graspTTA}. We visualize the curve of penetration - displacement by plotting the average displacements from samples with penetration volumes in $[0, 1\unit{\centi\metre^3})$, $[1\unit{\centi\metre^3}, 2\unit{\centi\metre^3})$, ..., $[12\unit{\centi\metre^3}, \infty)$ in Fig.~\ref{fig:disp-pene}.

It is obvious that penetrations and displacements are negatively correlated, which indicates the trade-off between the two criteria: more stable grasps are usually tighter and of more penetrations. Therefore, more penetrations does not necessarily lead to implausible grasps, and simulation displacements reflect stability better if compared under similar penetrations. In Fig.~\ref{fig:disp-pene}, our penetration curve is generally the lowest considering all penetration levels. Our displacements are especially low for small penetrations ($<$\SI{3}{\centi\metre^3}), which explains the performance improvement in distinction rate.

\begin{table}[t]
\setlength{\tabcolsep}{1mm}
\centering
\small{
\begin{tabular}{@{}lcccc@{}}
\toprule
\textbf{Methods} & \thead{Contact \\ Ratio (\%) $\uparrow$} & \thead{Penetration \\ Vol. (\unit{\centi\metre^3})$\downarrow$}  & \thead{Entropy} $\uparrow$ & \thead{Cluster\\ Size $\uparrow$} \\ \midrule
\multicolumn{5}{l}{\thead[l]{\hspace{-0.5em}\textbf{Results on GRAB dataset} \cite{GRAB:2020}\\ \vspace{-1.5em}}}\\
\midrule
GrabNet \shortcite{GRAB:2020}     & \textbf{100} & 8.50           & 2.80          & \textbf{4.40}           \\
ContactGen \shortcite{liu2023contactgen}      & 91.7          & 4.02          & 2.75          & 4.17          \\
FastGrasp \shortcite{wu2024fastgrasp}    & 98.3  & \textbf{3.03}   & 2.81 & 3.21 \\ \midrule
Ours      &  \textbf{100} & 5.13          & \textbf{2.83}          & 4.37 \\
\midrule \midrule
\multicolumn{5}{l}{\thead[l]{\hspace{-0.5em}\textbf{Results on out-of-domain HO3D dataset} \cite{hampali2020honnotate} \\ \vspace{-1.5em}}}\\ 
\midrule
GF  \shortcite{karunratanakul2020grasping}        & 73.0          & 9.68          & 2.85 & 2.16          \\
GrabNet  \shortcite{GRAB:2020}     & \textbf{100} & 12.89         & \textbf{2.88}          & 4.10          \\
GraspTTA  \shortcite{jiang2021graspTTA}    & \textbf{100} & 5.34          & 2.76          & 3.39          \\
ContactGen \shortcite{liu2023contactgen}    & 85.0          & \textbf{2.84} & 2.84          & \textbf{4.58} \\
FastGrasp \shortcite{wu2024fastgrasp}  & 92.5          & 4.79 & 2.81          & 3.30 \\ \midrule
Ours        & \textbf{100} & 6.62          & 2.80          & 4.47 \\ \bottomrule
\end{tabular}
}
\caption{Quantitative results of geometric plausibility and diversity criteria on both GRAB and HO3D Benchmarks. Note that although our method produces more penetrations, it is unavoidable due to more steady grasps.}\label{tab:penetration_diversity}
\end{table}

\subsection{Force Prediction Analysis}

We test the precision of force prediction by comparing the force values in contact regions with simulated results. While there is no ground truth for synthesized poses, we can obtain the force by running the labeling pipeline for it. We also construct an average predictor by predicting all forces as the average force from the training set as the baseline method. The results of our method vs. average predictor are \SI{0.21}{\newton} vs. \SI{0.51}{\newton} for point-level forces and \SI{4.13}{\newton} vs. \SI{17.6}{\newton}, which indicates that capturing the force distribution using a simple model like VAE is practical. Fig.~\ref{fig:force_pred} show a sample of force prediction, where we conclude that using average force can lead to very large force for all contacts but our method can judge where the force should be large and vice versa, thus producing results closer to the ground truth. 


\subsection{Ablation Study}

\begin{table}[t]
\centering
\setlength{\tabcolsep}{1.05mm}
\small{
\begin{tabular}{@{}lcccccc@{}}
\toprule
No. & \thead{Force \\ Branch} &  \thead{Key\\point} & \thead{Force\\value} & \thead{Simulation \\ Disp. (\unit{\centi\metre})$\downarrow$}& \thead{Penetration \\ Vol. (\unit{\centi\metre^3})$\downarrow$} & \thead{Distinction \\ Rate (\%)$\uparrow$} \\ \midrule
1                &    \ding{55}     &        \ding{55}           &     \ding{55}               &   2.26     & 7.27      &   5.0             \\
 2        &        \ding{51}            &      \ding{55}             &    \ding{55}       &   2.02     &  6.31   &   5.0              \\
 3  &      \ding{51}              &     \ding{55}    &        \ding{51}              & 1.60       &  7.70      &    6.7             \\
 4       &      \ding{51}             &     \ding{51}    &        \ding{55}                & 0.75      &  6.54       &   15.0              \\
5            & \ding{51}         &    \ding{51}         & \ding{51}                         & \textbf{ 0.68 } & \textbf{5.13} &  \textbf{ 17.5 }    \\ \bottomrule
\end{tabular}
}
\caption{Ablation study using GRAB. The numbers of test means: 1. No force prediction in training and inference; 2. Use force map in training, but not in generation; 3. Use force map only as weights in optimization 4. Replace force predictions with average value; 5. Our full method. } \label{tab:ablation}
\end{table}


\begin{table}
\centering
\setlength{\tabcolsep}{1.1mm}
\small{
\begin{tabular}{cccccccc}
\toprule
$n_{kp}$ & 0    & 1    & 2    & 3    & 4    & 5    & 16   \\ \midrule
Simu. Disp. (\unit{\centi\metre})$\downarrow$& 2.02 & 1.78 & 1.48 & 0.68 & 0.66 & 0.67 & \textbf{0.56} \\
Pene. Vol. (\unit{\centi\metre^3})$\downarrow$ & 6.31 & \textbf{5.00} & 5.29 & 5.13 & 7.95 & 8.49 & 9.02 \\
Distinction Rate (\%)$\uparrow$ & 5.0  & 10.0 & 7.5  & \textbf{17.5} & 6.7  & 7.5  & 5.8  \\ \bottomrule
\end{tabular}
}
\caption{Ablation study on the number of keypoints $n_{kp}$ using GRAB. $n_{kp}=3$ is chosen for its highest distinction rate.}\label{tab:ablation_nkps}
\end{table}

In this section, we first study the effect of force prediction. We set up 5 different experiment configurations and report the settings and results in Tab.~\ref{tab:ablation}. Generally, using keypoint-based optimization and integrating exact force values are the components with the largest contributions. By comparing the increase from 2 to 3 and from 2 to 5, we can conclude that using force simply as weights is not as effective as using keypoints. The obvious increase from 2 to 4 demonstrates the effectiveness of keypoints in optimization. The last two rows show that using predicted force values effectively reduces unreasonable penetrations thus improving distinction rate. It indicates the predicted forces can mimic real forces and help to make keypoints easier to fit.

Moreover, we conduct an ablation to study the optimal number of keypoints and report the result in Tab.~\ref{tab:ablation_nkps}. The trend is clear: with the increase of keypoint number, the displacement decreases and the penetration increases. The distinction rate, considering both stability and penetrations, achieves the peak value at $n_{kp}=3$. Essentially, although more keypoints lead to smaller displacements, it also comes with a higher risk of unreachable combinations, thus causing both inter- and self-penetrations. Therefore, we choose $n_{kp}=3$ as the optimal keypoint number.


%% file: contents/conclusion.tex
\section{Conclusion}

In this paper, we propose a novel contact-based grasp generation pipeline to improve stability. We extend the contact representation to include contact forces and derive an energy function to model stability by accelerations. Then, stability-aware key contact points are identified using the energy to guide the pose generation. The method achieves state-of-the-art stability in grasp synthesis, suggesting the significance of modeling contact forces and keypoint-guided optimization.



%% file: contents/acknowledgement.tex
\section{Acknowledgements}

This research was funded by the China Scholarship Council - University of Birmingham PhD Scholarship programme (No. 202406230091), the MSIT(Ministry of Science and ICT), Korea, under the ITRC(Information Technology Research Center) support program(IITP-2025-RS-2020-II201789) supervised by the IITP(Institute for Information \& Communications Technology Planning \& Evaluation), and National Natural Science Foundation of China (Grant No. 62406163).

We would like to thank Yuming Chen for discussions on simulation strategies. We also thank Jungchan Cho and Jeongho Lee for their suggestions on paper writing.